
\input phyzzx

     \FRONTPAGE
\line{\hfill BROWN-HET-974}
\line{\hfill IFUP-TH75/94}
\line{\hfill December 1994}
\def\pv{-\hskip-13pt\int}
\vskip1.0truein
\titlestyle{{SIMPLICIAL CHIRAL MODELS
}\foot{Work supported in
part by the Department of Energy under
contract DE-FG02-91ER40688-Task A}\break}
\vskip .25in

\author{Paolo ROSSI}
\centerline{{\it Dipartimento di Fisica dell'Universit\'a}}
\centerline{{\it and I.N.F.N., I-56126, Pisa, Italy}}

\centerline{and}
\author{Chung-I TAN}
\centerline{{\it Department of Physics}}
\centerline{{\it Brown University, Providence, RI 02912, USA}}
\bigskip
\abstract
Principal chiral models on a d-1 dimensional simplex are introduced
and studied analytically in the large $N$ limit.  The $d = 0 ,
2, 4$ and $\infty$ models are explicitly solved.  Relationship with
standard lattice models and with few-matrix systems in the double
scaling limit are discussed.

\endpage

The importance of understanding the large $N$ limit of matrix-valued field
models cannot be overestimated.  Not only is this the basic ingredient of the
$1/N$ expansion in the physically relevant case of QCD, but also the existence
of large $N$ criticalities for finite values of the coupling is the starting
point for the approach to 2-dimension quantum gravity known as the double
scaling limit.  Moreover we note that solutions of few-matrix systems may have
a
direct application to more complex systems in the context of strong-coupling
expansion, since they may be reinterpreted as generating functionals for
classes
of group integrals that are required in strong coupling calculations [1].
Unfortunately our knowledge of exact solutions for the large $N$ limit of
unitary matrix models is still limited.  After Gross and Witten's
solution [2] of the single-link problem, exact results were obtained only for
the external field problem [3,4] and a few toy models
$(L=3,4$ chiral chains) [5,6].

In this letter we introduce a new class of lattice chiral models, whose large
$N $ behavior can be
analyzed by solving an integral equation for the eigenvalue distribution of a
single hermitian semi definite positive matrix. The models we are going to
study
are principal chiral models, with a global $U(N) \times U(N)$ symmetry, defined
on a (d-1)-dimensional simplex formed by connecting in a fully symmetric way
$d$
vertices by ${(d-1)(d-2)\over 2} $ links.   The partition function for such a
system is defined to be
$$
Z_d= \int \, \prod_{i=1}^d \, dU_i \, \exp \, N\beta \sum_{i>j=1}^d \, \Tr
\,[U_i U_j^{\dagger} + U_j U_i^{\dagger} ]. \eqno\eq
$$
 Despite its apparent simplicity, this class of models
includes most of the previously known solvable systems. As a function of the
parameter $d$, which specifies the coordination number of lattice sites, it
interpolates between the two-dimensional Gross-Witten model (with third order
phase transition) and the infinite-dimensional mean field solution (showing a
first order phase transition) of standard infinite volume lattice models. It
also includes a case which, in the double scaling limit, corresponds to that of
a $c=1$ conformal field theory.

It is possible to eliminate the direct interactions among the unitary matrices
$U_i$ by introducing an identity in the form
$$1 = {\int dA \, \exp \, - N\beta \,Tr\, [A - \sum_{i=1}^d U_i ] [A^{\dagger}-
\sum_{j=1}^d U_j^{\dagger} ]\over
\int dA \exp - N\beta \, \Tr AA^{\dagger}},\eqno\eq
$$
where $A$ is a $N\times N$ complex matrix.  As a consequence we obtain $Z_d =
\tilde{Z}_d/\tilde{Z}_0$ where
$$\tilde{Z}_d = \int \prod_{i=1}^d dU_i \, dA \, \exp \left\{ - N\beta \,Tr \,
AA^{\dagger} + N\beta\, Tr
\, A \sum_i \, U_i^{\dagger} + N\beta\, Tr \, A^{\dagger} \sum_i U_i - N^2
\beta
d\right\}.\eqno\eq
$$
We now introduce the function
$$F[BB^{\dagger} ] = {1\over N^2} \, \log \int dU \exp {N\over 2} Tr
\,[BU^{\dagger} + UB^{\dagger} ].\eqno\eq
$$
$F$ is a known function of the eigenvalues $x_i$ of the hermitian semipositive
definite matrix
$BB^{\dagger}$.  More specifically, in the large $N$ limit, we know that [3,4]
$$
F(x_i ) = {1\over N} \sum_i (r + x_i )^{1/2} - {1\over 2N^2} \sum_{i,j} \log
\left[ {(r + x_i )^{1/2} + (r + x_j )^{1/2}\over 2} \right] - {r\over 4} -
{3\over 4},  \eqno\eq
$$
and there are two distinct phases:\nextline
a) weak coupling $\quad ~~ r = 0 $,\nextline
b) strong coupling $\quad{1\over N} \sum_i (r + x_i)^{-1/2} = 1$.\nextline
Up to irrelevant factors, it follows that
$$
\tilde{Z}_d = \int dB \, \exp \, \left\{ - {N\over 4\beta} \, Tr \,
BB^{\dagger} + N^2 d F (BB^{\dagger} ) - N^2
\beta d\right\}, \eqno\eq
$$
where $B$ replaces $2\beta A$.

Morris [7] has shown that the angular integration can be performed in the case
of complex matrices, and, again up to irrelevant factors, we may replace eq.
(6)
by
$$
\tilde{Z}_d = \int_0^{\infty} dx_i \, \prod_{i\not= j}  (x_i - x_j ) \exp
\left\{\, - {N\over 4\beta}
\sum_i x_i + N^2 d\, F(x_i ) - N^2 \beta \right\}.\eqno\eq
$$
In the large $N$ limit it is legitimate to evaluate this integral by a saddle
point method.  The saddle-point equation resulting from eqs. (5) and (7) is
$$
{\sqrt{r+x_{i}}\over 2\beta} - d = {1\over N} \sum_{j\not= i} \, {(4-d)\sqrt{r
+
x_i} + d\sqrt{r+x_j}\over x_i - x_j }.\eqno\eq
$$

We introduce a new variable $ z_i \equiv \sqrt{r + x_i}$ and, in the large $N$
limit, we assume that these eigenvalues lie in a single interval $[a,b]$.
Denoting the large $N$ eigenvalue density by
$\rho (z)$, eq. (8) becomes an integral equation for $\rho(z)$,
$${z\over 2\beta } - d = \pv_a^b \, dz' \rho (z') \left[ {2\over z - z'} -
{(d-2)\over z + z'}
\right], \eqno\eq
$$
where the integration region is restricted by the condition, $0\leq a\leq b$,
with   $a$ and
$b$  determined dynamically.  In particular, the normalization condition,
$$\int_a^b \rho (z') dz' = 1, \eqno\eq
$$
must always be satisfied. Furthermore, one has the following constraint,
$$
\int_a^b \, \rho (z') {dz'\over z'}\leq 1,  \eqno\eq
$$
with the equality holding exactly in the strong coupling region where
$a=\sqrt{r}$.

Let us begin by first discussing several simple cases where eq. (9) can be
solved readily. For
$d=0$, the problem reduces to one with a pure gaussian interaction, and, by a
more or less standard technique, one finds that
$$
\rho (z) = {1\over 4\pi\beta} \, \sqrt{16\beta - z^2}.\eqno\eq
$$
For $d=0$,  there is no weak coupling phase. Note also that, since the $F$-term
in eq. (6) vanishes for $d=0$,  eq. (12) is  obtained with
$a=0$. As a consequence, up to a constant,
$$
\tilde{Z}_0 = \exp \, N^2 \ln \beta, \eqno\eq
$$
as expected.

When $d=2$ we obtain
$$
\eqalign{ \rho_w (z) & = {1\over 4\pi\beta} \sqrt{8\beta - (z -4\beta )^2}
\qquad\qquad\qquad\qquad\qquad\qquad \beta\geq {1\over 2},\cr
\rho_s (z) & = {1\over 4\pi\beta} z {\sqrt{(1+6\beta) -z\over z-(1-2\beta)}},
\quad \quad r(\beta ) = (1-2\beta )^2, \quad\quad\,   \beta \leq {1\over 2},
}\eqno\eq
$$
and one may show that all results are consistent with a reinterpretation of the
model as a Gross-Witten [2] one-plaquette model, with $\beta_c = {1\over 2}$.

When $d=3$ the model can be mapped into the three-link chiral chain, which is
known to possess a third order phase transition at $\beta_c = {1\over 3}$ [5].

The first non-trivial and new situation begins at $d=4$. We have explicitly
solved  the $d=4$ model, both in the weak and in the strong coupling phase.
The
eigenvalue density may be expressed in terms of elliptic integrals, and
supplementary conditions allow for the determination of $a$ and
$b$.

Introducing the variable $k(\beta ) =\sqrt{ 1 - {a^2\over b^2}}$, in weak
coupling we obtain, in terms of standard elliptic integrals $K$, $\Pi$, and
$E$,
$$
\rho_w (z) = {b\over 2\pi^2 \beta} \, (z^2 - a^2 )^{1/2} (b^2 - z^2 )^{-1/2}
\left[ K (k) - {z^2\over b^2 } \,\Pi\, (1 - {z^2\over b^2} , k)\right],\eqno\eq
$$
with the condition
$4\pi\beta = bE(k)$.
In strong coupling we have
$$
\rho_s (z) = {1\over 2\pi^2\beta}\, {z^2\over b} (z^2 - a^2 )^{-1/2}\, (b^2 -
z^2 )^{-1/2} \left[ (b^2 - a^2 ) K (k) - (z^2 - a^2 ) \Pi (1 - {z^2\over b^2} ,
k )\right],\eqno\eq
$$
with the condition
$4\pi \beta = b [ E(k) - {a^2\over b^2} K (k) ]$.
In both regimes, eq. (10) must also be satisfied.
Closed form solutions for the constraints may be obtained at criticality: when
$\beta = \beta_c = {1\over 4}$, we get $a=0 ,\quad  b = \pi$, and
$$
\rho_c (z) = {z\over \pi^2} \log \, {1 + \sqrt{1 - {z^2\over \pi^2}}\over 1 -
\sqrt{1 - {z^2\over
\pi^2}}}.\eqno\eq
$$

Let's finally observe that a large-$d$ solution of eq. (9) may easily be found
by
assuming $\rho (z)\rightarrow \delta (z - \bar{z})$.  The weak coupling
solution is
$$
\bar{z} = \beta d \left[ 1 + \sqrt{1 - {1\over \beta d}}\right], \quad\quad
\beta \geq \beta_c =
{1\over d}, \eqno\eq$$ and for strong coupling $\bar{z} = 0$.  Amusingly
enough, this solution turns
out to coincide with the large $D\equiv {d\over 2}$ mean field solution [8] of
infinite volume
principal chiral models on D-dimensional hypercubic lattices with the same
coordination number as our
corresponding models.

We would like to add a few comments. Solving eq. (9) is certainly a well
defined problem for any
value of $d$, and in particular we expect to be able to find explicit solutions
for simple cases,
like $d=1$ and $d=3$.  It is also possible to analyze eq. (9) numerically;
details of our analytical
and numerical techniques will be reported elsewhere; we only mention that for
sufficiently large
$\beta > \beta_c$ we can get the eigenvalue distribution with desired accuracy,
while near
criticality convergence is slow:  however within 1\% accuracy we have evidence
that $\beta_c =
{1\over d}$ for all integer values of $d$ [9].
It would be quite interesting to achieve more information, both qualitative and
quantitative, on
the $d$-dependence of the phase transition.

The thermodynamical quantity whose computation is easiest is the internal
energy per unit link,
$w_1$, which may be obtained from
$$
d (d-1) w_1 = {1\over 4\beta^2} \int_a^b dz \rho (z) (z^2 - r ) - d - {1\over
\beta}.\eqno\eq
$$
One may then extract, in the vicinity of $\beta_c$, the critical exponent for
the specific
heat,
$\alpha$.
At present we know that when $d=2$, $  \alpha = -1$, when $d=3$, $\alpha = -
{1\over 2}$,
when
$d = 4$, $\alpha = 0$, and for sufficiently large $d$ the transition is
first-order, {\it  i.e.},
$\alpha=1$.

It is worth observing in this context that a more general model involving four
unitary matrices and
three couplings, interpolating between our $d=4$ case and the 4-link chiral
chain, can be
re-expressed as a model of two coupled complex matrices and admits many
solvable limits, all
characterized by $\alpha =0$, which corresponds to a $c=1$ conformal field
theory.
\vskip .10in
\noindent{\bf References}
\vskip .10in
\pointbegin
M. Campostrini, P. Rossi and E. Vicari, Pisa preprrint IFUP-TH63/94.
\point
D. J. Gross and E. Witten, {\it Phys. Rev.} {\bf D 21},446 (1980).
\point
R. C. Brower and M.Nauenberg, {\it Nucl. Phys.} {\bf B180}, 221 (1981).
\point
E. Brezin and D. J. Gross, {\it Phys. Lett.} {\bf 97B}, 120 (1980).
\point
R. C. Brower, P. Rossi and C-I Tan, {\it Phys. Rev.} {\bf D23}, 942, 953
(1981).
\point
D. Friedan, {\it Comm. Math. Phys.} {\bf 78}, 353 (1981).
\point
T. R.
Morris, {\it Nucl. Phys.} {\bf B356}, 703 (1991).
\point
J. B. Kogut, M. Snow and M. Stone, {\it Nucl. Phys.} {\bf B200}, 211 (1982).
\point
M. Campostrini and P. Rossi (unpublished).

\bye